\documentclass[twocolumn,showpacs,preprintnumbers,amsmath,amssymb]{revtex4}
\makeatletter
\def\@dotsep{4.5}
\makeatother
\usepackage[dvips]{graphicx}  \usepackage{amsmath}
\usepackage{color}
\definecolor{textcolor}{cmyk}{0,0,0,1}
\definecolor{magenta}{rgb}{1,0,1}
\definecolor{green}{rgb}{0,1,0}
\definecolor{red}{rgb}{1,0,0}

\begin{document}
\draft
\title{ Carbon nanoelectronics: unzipping tubes into graphene ribbons}
\author{ H. Santos, L. Chico and L. Brey}

\affiliation{
Instituto de Ciencia de Materiales de Madrid, Consejo Superior de
Investigaciones Cient{\'{\i}}ficas, Cantoblanco, 28049 Madrid, Spain}

\date{\today}

\begin{abstract}
We report on the transport properties of novel carbon nanostructures
made of partially unzipped carbon nanotubes, which can be regarded
as a seamless junction of a tube and a nanoribbon.  We find that
graphene nanoribbons act at certain energy ranges as a perfect
valley filters for carbon nanotubes, with the maximum possible
conductance. Our results show that a partially unzipped carbon
nanotube is a magnetoresistive device, with a very large value of
the magnetoresistance. We explore the properties of several
structures combining nanotubes and graphene nanoribbons,
demonstrating that they behave as optimal contacts for each other,
and opening a new route for the design of mixed graphene/nanotube
devices.
\end{abstract}
\pacs{81.05.Uw, 73.20.-r, 85.75.-d}

\maketitle

\emph{Introduction.} The experimental isolation of
graphene \cite{Novoselov_2004} and the anomalous electronic
properties of its carriers \cite{Novoselov_2005,Zhang_2005} has
rapidly motivated intense theoretical and experimental
investigation,
among many other characteristics, of its transport properties
\cite{Castro_Neto_RMP}. The valence and
conduction bands of graphene touch at two inequivalent points of the Brillouin
zone. Near these points the dispersion relation is linear, so graphene
carriers behave as massless Dirac fermions. The large separation in
reciprocal space of the Dirac points suppress intervalley
scattering in pure graphene samples \cite{Morpurgo_2006}. Thus,
besides the spin and charge degree of freedom, graphene
carriers should be also characterized by a valley
index \cite{Castro_Neto_RMP,Rycerz_2007}.

Due to its two-dimensional character, graphene can be patterned
using high-resolution lithography \cite{Berger_2006}, so in
principle nanocircuits with transistors and interconnects can be
fabricated in the same graphene layer in a fully compatible way with
the present electronic technology. In these nanodevices graphene
nanoribbons could be used as connectors \cite{Iyengar_2008}.
Lithographic techniques have been employed to produce wide ($>$20
nm) stripes of graphene \cite{Han_2007,Ozyilmaz_2007}, but with
limited smoothness due to limitations in the resolution. Chemical
\cite{Schniepp_2006,Li_2008} and synthetic \cite{Yang_2008} methods
have been employed successfully, albeit producing microscopic
quantities of graphene nanoribbons. Bulk production of ribbons has
been achieved with a chemical vapor deposition method, but the
samples had a wide dispersion in size and number of layers
\cite{Campos-Delgado_2008}, so the controlled fabrication of
nanoribbons of small width remained as a technological challenge.

Quite recently, two experimental groups announced simultaneously a
promising way to fabricate narrow graphene nanoribbons (GNR) using
carbon nanotubes as starting material
\cite{Kosynkin_2009,Jiao_2009}. These two groups propose to
longitudinally unzip carbon nanotubes (CNTs) to make nanoribbons,
either by chemical attack \cite{Kosynkin_2009} or by plasma etching
\cite{Jiao_2009}, with very high yields. Unzipping carbon nanotubes
appears as a promising way to fabricate narrow nanoribbons needed
for nanoelectronic applications.

In this Letter we report on the electronic transport properties of
unzipped carbon nanotubes. We propose that partial unzipping of
carbon nanotubes can actually be used to produce a new class of
carbon-based nanostructures, which combine nanoribbons and
nanotubes. By studying the GNR/CNT junction we conclude that
nanoribbons and nanotubes behave as ideal contacts for each other.
Furthermore, we obtain that structures formed by zigzag-terminated
GNR and armchair CNT units behave as spin and valley filters, and
can be used as building blocks for carbon-based devices featuring very large
magnetoresistance.

\begin{figure}[h]
\includegraphics[width=\columnwidth,clip]{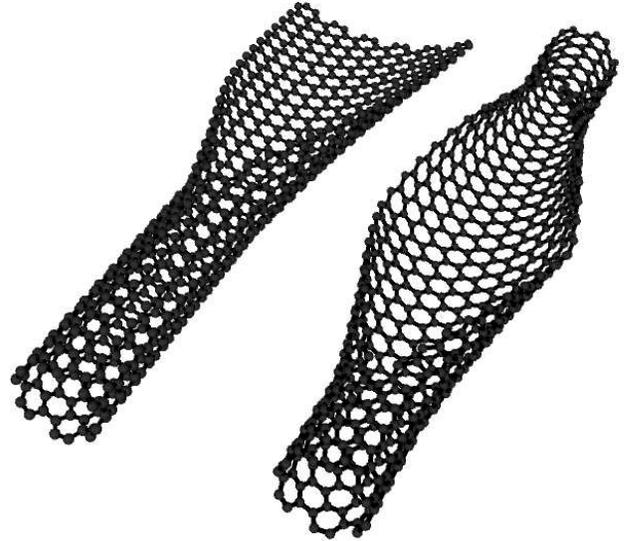}
\caption{ (Color online)
Geometry of two partially unzipped nanotubes. Left: a (6,6) armchair nanotube unzipped into a 12-ZGNR, making a CNT/GNR single junction. Right: the same nanotube unzipped in its central part, yielding a zigzag nanoribbon quantum dot connected to armchair nanotube contacts.}
\label{fig:fig1}
\end{figure}

\begin{figure}
\includegraphics[width=\columnwidth,clip]{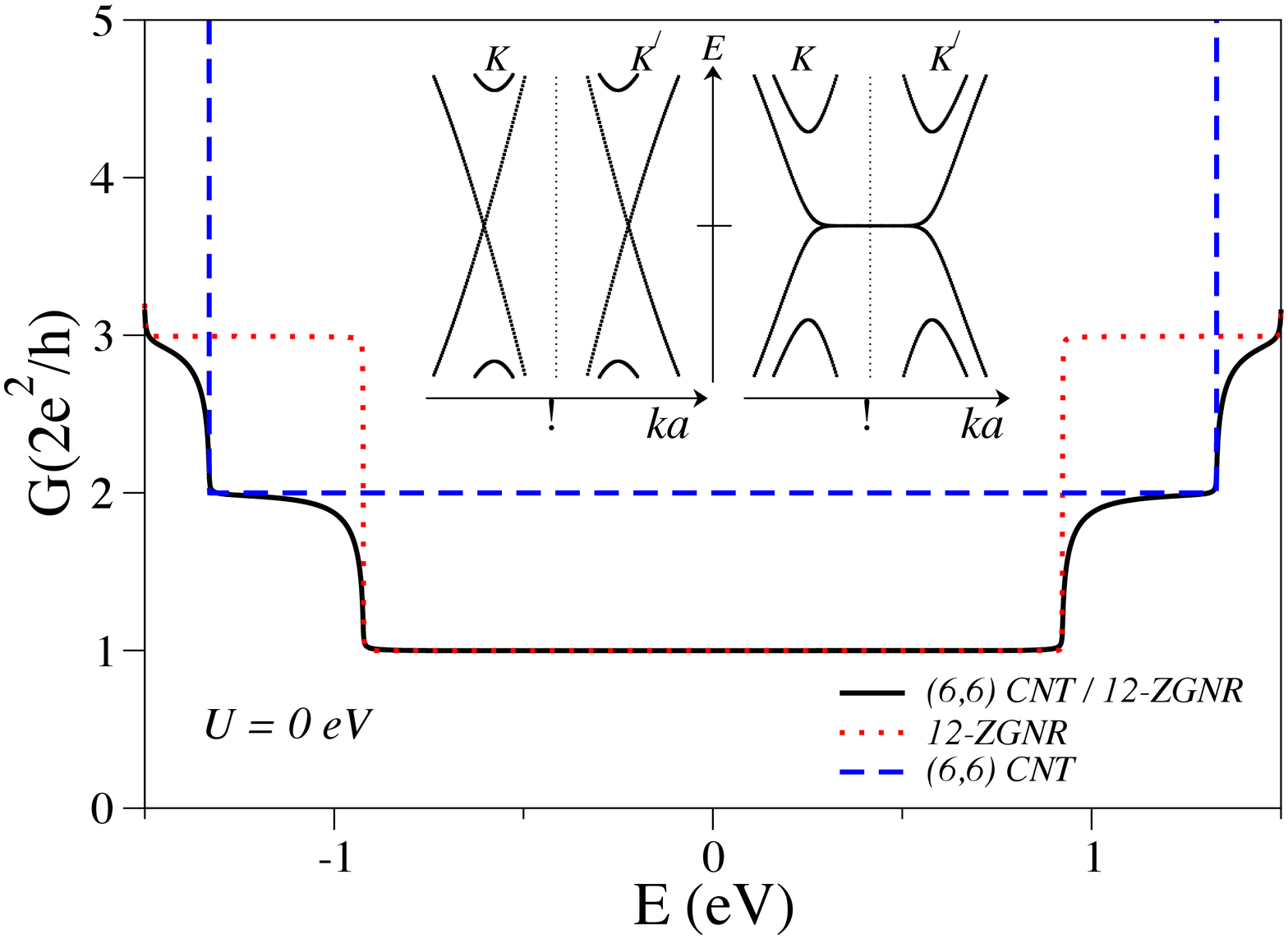}
\caption{ (Color online) Solid line: conductance of a (6,6)
CNT/12-ZGNR junction. Dashed (dotted) line: conductance of a perfect
infinite (6,6) CNT (12-ZGNR). The results correspond to
$U$=0. The insets show the bandstructures of the armchair
nanotube (left) and the zigzag nanoribbon (right). The flat bands
joining the Dirac points in the ZGNR corresponds to states localized
at the edges of the ribbon.} \label{fig:figprop}
\end{figure}

\noindent \emph{Ingredients.}
\par \noindent
i) Carbon nanotubes are rolled-up cylinders of
graphene \cite{Iijima_1991}. Their electronic properties can be
approximately derived from the graphene band structure by imposing
the Born-von Karman boundary condition \cite{Saito_book}, and
depending on their geometry, can behave as metals or semiconductors
\cite{Hamada_1992,DresselhausPW_1996}. In this work, we focus in
armchair carbon nanotubes that correspond to the configuration with
no 'twist' in the graphene rolling, with an armchair shape of the
cross section along the circumference \cite{Saito_book}. Armchair
nanotubes are denoted by $(n,n)$, being 2$n$ the number of carbon atoms at
the CNT circumference; they are one-dimensional metals with
two inequivalent Fermi points in the Brillouin zone, reminiscent of the graphene Dirac
points (See inset of Fig. \ref{fig:figprop}).
\par \noindent
ii) Carbon nanoribbons are obtained by cutting graphene in the form
of a quasi-one-dimensional stripe. The electronic properties of GNR strongly
depend on the atomic edge termination. There are two basic
shapes for graphene edges, armchair and zigzag \cite{Fujita_1996}.
The GNR electronic properties can be derived from the graphene band
structure by imposing the appropriate boundary
conditions \cite{Brey_2006b}. Armchair GNR can be either metallic or
semiconducting depending of their width, whereas for the zigzag GNR (ZGNR)
two-fold degenerated flat bands lie at the Fermi energy. These bands
are associated with edge states \cite{Brey_2006b} and their
dispersionless character favors an insulating antiferromagnetic ground state,
with opposite magnetization at the
edges \cite{Fujita_1996,JFR_2008,Son_2006a,Pisani_2007,Munoz-Rojas_2009}.
Zigzag graphene nanoribbons are obtained by unrolling an
armchair CNT, see Fig. \ref{fig:fig1}.  The ZGNR width
is defined by the number $n$ of zigzag rows in the unit cell; the usual notation for
such ribbon is $n$-ZGNR.

In this work we study partially unzipped carbon nanotubes, such as
those shown in Fig. \ref{fig:fig1}, which are equivalent to a combination of carbon
nanotubes and graphene nanoribbons. In what follows, we focus in
armchair carbon nanotubes and the derived graphene zigzag
nanoribbons.

\noindent \emph{Model Hamiltonian and transport calculation.} We
describe the motion of carriers between carbon $\pi$
orbitals with a single first-neighbor hopping parameter $t$. The
electronic repulsion is included within the Hubbard model, which we
solve in the mean field approximation. For $t\simeq 3$ eV and the
Hubbard term  in the range 1.5 eV $ \! < \!  U \!  < \!  $ 3 eV, this
approach describes adequately the main features of the \emph{ab
initio} calculations \cite{JFR_2007,JFR_2008} around the Fermi energy
of graphene nanoribbons. The Hamiltonian reads
\begin{equation}
H = - t \sum_{i,j,\sigma}  c^\dagger_{i,\sigma} c_{j,\sigma}+ U
\sum_{i,\sigma} n_{i,\sigma} < n_{i,-\sigma} > ,
\label{eq:ham}
\end{equation}
where $ c^\dagger_{i,\sigma}$ ($ c_{j,\sigma}$) is a creation
(annihilation) operator at atom $i$ ($j$) of a $\pi$ electron with
spin $\sigma$. We consider that the effect of the unzipping is to
cut the hopping between the carbon atoms where the opening occurs,
 and we assume that unzipping does not modify the hopping parameter
between the other carbon atoms, which we set to $t=2.66$ eV. This Hamiltonian depends on the electronic
occupation and has to be solved selfconsistently \cite{LopezSancho_2001}.

As the systems lack translational invariance, we follow a Green
function (GF) approach to calculate the electronic and transport
properties \cite{Chico_1996b,Nardelli_1999}. To this purpose, we
divide the system into three parts, namely a central region
connected to the right and left leads. The Hamiltonian can thus be
written as
\begin{equation}
H= H_C  + H_R + H_L + h_{LC} + h_{LR},
\end{equation}
where $H_C$, $H_L$, and $H_R$ are the Hamiltonians of the central
portion, left and right leads respectively, and $h_{LC}$, $h_{RC}$
are the hoppings from the left $L$ and right $R$ lead to the central
region $C$. The GF of the latter is
\begin{equation}
\mathcal{G}_C(E) = (E-H_C - \Sigma_L -\Sigma_R)^{-1},
\end{equation}
where $\Sigma_\ell= h_{\ell C}g_\ell h_{\ell C}^\dagger$ is the
selfenergy due to lead $\ell=L,R$, and $g_\ell = (E -H_\ell)^{-1}$
is the GF of the semiinfinite lead $\ell$.

In a zero bias approximation, the Landauer conductance $G$ is given
by
\begin{equation}
G = \frac{2e^2}{h}\mathcal{T}(E_F).
\end{equation}
$\mathcal{T}$ is the transmission function, that can be obtained
from the GFs of the central part and the leads:
\begin{equation}
\mathcal{T}(E)  = {\rm Tr} (\Gamma_L(E) \mathcal{G}_C (E)\Gamma_R(E)
\mathcal{G}_C^\dagger(E)),
\label{eq:transm}
\end{equation}
where $\Gamma_\ell=i(\Sigma_\ell -\Sigma_\ell^\dagger)$ describes
the coupling of the central region $C$ to lead $\ell$. For a single junction, the left and right
electrodes are connected directly, so $\mathcal{G}_C$ in Eq. \ref{eq:transm} is just the GF of the interface
between the left and right lead, $\mathcal{G}_{LR}$.

\emph{Results.} In Fig. \ref{fig:figprop} we show the conductance of
a single (6,6) CNT/12-ZGNR junction, as the one depicted in the
upper panel of Fig. \ref{fig:fig1}, for noninteracting electrons
($U=0$). As a reference, we show in dashed and dotted lines the
conductances of the perfect infinite carbon nanotube and nanoribbon.
Around zero energy, the conductance of the junction is equal to that
of the perfect nanoribbon, demonstrating that the (6,6) CNT acts as
a transparent contact for the 12-ZGNR. Assuming left to right
conduction, the ZGNR only has the $K'$ channel open to transport and
it is completely transparent to states from the corresponding $K'$
valley of the CNT. This indicates that
backscattering is practically zero in the device and the conductance
in this energy range is set by the ribbon, which acts as a valley
filter for the carbon nanotube.
Our numerical results show that states from the $K$ valley of the
CNT cannot transverse the junction.   At higher energies, more channels
open in the ribbon, so the conductance now is limited by that of the
(6,6) nanotube: here, the ribbon is acting as a transparent contact
for the carbon nanotube, and transport from both valleys is allowed.
Calculations with larger systems, such as a (18,18) CNT/36-ZGNR (not
shown here) yield similar results, with the obvious increase in the
number of channels at energies closer to the Fermi level. We have
also investigated the narrowing of the ribbon part, such as a (6,6)
CNT/10-ZGNR, or a (18,18)/32-ZGNR, finding that the effect is robust
against the precise form of the CNT/ZGNR junction.

\begin{figure}
\includegraphics[width=\columnwidth,clip]{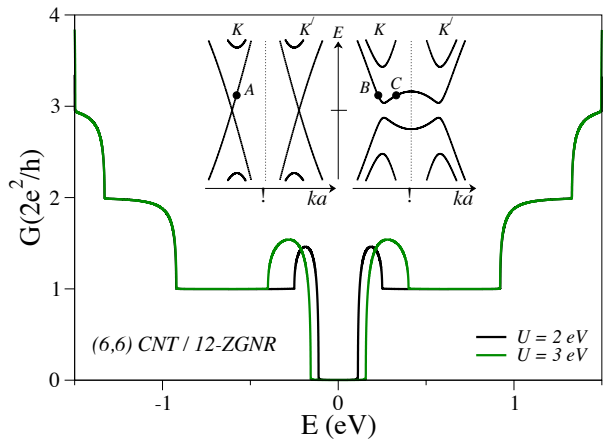}
\caption{ (Color online) Conductance of a (6,6) CNT/12 ZGNR with
on-site repulsion ($U=2$ and 3 eV).  Insets show the bandstructures of the armchair
nanotube (left) and the zigzag nanoribbon (right) for $U$=3 eV.}
\label{fig:fig3}
\end{figure}

As discussed above, the interaction between electrons changes
dramatically the band structure of graphene nanoribbons near the
Fermi energy. In the following we analyze how the
 interactions
modify the low energy transport properties. Fig. \ref{fig:fig3} shows
the effect of electron-electron interactions in the conductance of
the (6,6)CNT/12-ZGNR junction for $U=2$ eV and $U=3$ eV. The inset
shows the  bandstructure  for the (6,6)CNT and for the 12-ZGNR for
an interaction $U=3$ eV.  For
CNTs the Hubbard interaction does not modify the magnetic moments
on the carbon atoms, so the effect of $U$  is just a rigid shift of the
electronic structure. In the case of ZGNR the interaction orders
ferromagnetically the most external atoms at each edge, and the magnetic
moments on opposite edges couple antiferromagnetically.
Magnetic order induces dispersion of the edge bands along the edge direction,
opening a gap at the center of the band
structure that increases with the value of the interaction $U$. The
states with opposite spin orientation are degenerated, but the edge
bands with opposite spin correspond to states located at opposite
edges. The many body-induced gap in ZGNR precludes transport
near the Fermi energy.
Above the gap, there is a region of enhanced conductance with
respect to the non-interacting case. This occurs because the
dispersion of the edge states,  induced by the the electron electron
interaction, opens a new electronic channel near each Dirac point of
the ZGNR, as for example at point C in the inset of Fig.
\ref{fig:fig3}. Besides transmission from $K'$ to $K'$ valleys
observed in the non-interacting case, now a state A in the CNT $K$
valley can be transmitted into state $C$ of the ZGNR at the same
valley, giving an enhanced conductance. The width of this bump in
the conductance is proportional to the midband gap and increases
with $U$. The state C is a edge state and for opposite spin
orientations the wavefunction of this state is localized in opposite
edges.  Therefore the excess of current  with respect to the
noninteracting case (bump regions of Fig. \ref{fig:fig3}) is
localized at the edges and with opposite spin polarization. Above
this energy region there is an energy interval where the conductance
gets the value $2e^2/h$, and valley filtering occurs, as explained
for the non-interacting case. This energy interval is around 0.6 eV
for $U=2$ eV, making it possible the observation of this filtering
in carbon nanotubes. Above this interval, the conductance values are
quite similar to the non-interacting cases, demonstrating the high
transparency of armchair CNTs for ZGNR and viceversa. The
application of a magnetic field makes the ZGNR ferromagnetic and
metallic \cite{Munoz-Rojas_2009}. The magnetic field  closes the
midgap, opening new channels at low energies and giving rise to a
large magnetoresistance. We propose that a single unzipped carbon
nanotube is by itself a device featuring 100 \% magnetoresistance.

\begin{figure}[h]
\includegraphics[width=\columnwidth,clip]{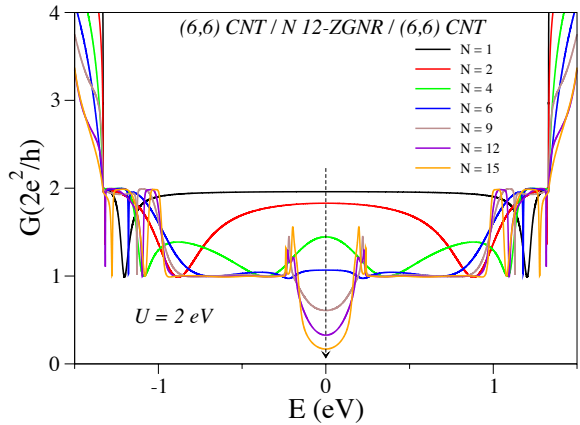}
\caption{(Color online) Conductance of a (6,6) CNT/N12-ZGNR/(6,6)CNT with
on-site repulsion $U=2$ eV for several ribbon lengths (N=1 to N=15). The arrow indicates increasing system size.} \label{fig:fig4}
\end{figure}

We have also explored the properties of double junction systems,
such as the one depicted in Fig. \ref{fig:fig1} (right), an infinite
armchair (6,6) nanotube open in its central part making a ZGNR. We
denote this structure as (6,6) CNT/N 12-ZGNR/(6,6) CNT, where N is
the number of unit cells in the nanoribbon. The transparency of the
nanotube contacts is evident in Fig. \ref{fig:fig4}: the
transmission through the central ribbon part is higher that in an
infinite ribbon for the smaller sizes, and slowly decays to the zero limit value in
the gap with increasing ribbon size. Other combinations, such a CNT
quantum dot with ribbon contacts can be envisioned, expanding the
possibilities of carbon electronics, in analogy to the quantum dot and superlattice structures
proposed for carbon nanotubes \cite{Chico_1998,Jaskolski_2005}.

In summary, we propose a new class of carbon nanostructures based on
unzipped nanotubes, which actually consist of mixed carbon
nanotube/nanoribbon systems. We have found that  ribbons from
unzipped tubes behave as completely transparent contacts  for the
parent tubes, and viceversa.  Our results demonstrate that partially
unzipped carbon nanotubes are by themselves magnetoresistive
devices, with a large value of the magnetoresistance. Furthermore,
carbon nanoribbons act as valley filters for carbon nanotubes; this
behavior is robust with respect to the inclusion of
electron-electron interaction, opening the possibility of exploiting
the valley degree of freedom in a new class of carbon-based
nanodevices.

This work has been partially supported by the Spanish DGES under
grants MAT2006-06242 and MAT2006-03741 and Spanish CSIC under grant
PI 200860I048.


\end{document}